\documentclass[twocolumn,aps,prl,showpacs,amsmath,amssymb,floatfix]{revtex4-1}
\usepackage{graphicx}
\usepackage{dcolumn}
\usepackage{bm}

\begin{document}

\title{Energy Loss at Propagating Jamming Fronts in Granular Gas Clusters}

\author{Justin C. Burton}
\email{Author to whom correspondence should be addressed:\newline justin.c.burton@emory.edu}
\affiliation{James Franck Institute, Enrico Fermi Institute and Department of Physics, The University of Chicago}
\author{Peter Y. Lu}
\author{Sidney R. Nagel}
\affiliation{James Franck Institute, Enrico Fermi Institute and Department of Physics, The University of Chicago}

\date{\today}

\begin{abstract} 
We explore the initial moments of impact between two dense granular clusters in a two-dimensional geometry. The particles are composed of solid CO$_{2}$ and are levitated on a hot surface. Upon collision, the propagation of a dynamic ``jamming front" produces a distinct regime for energy dissipation in a granular gas in which the translational kinetic energy decreases by over 90\%.  Experiments and associated simulations show that the initial loss of kinetic energy obeys a power law in time, $\Delta E=-Kt^{3/2}$, a form that can be predicted from kinetic arguments.
\end{abstract}


\maketitle

From the patterning of sand dunes \cite{Bagnold1954} to the intermittency of avalanches \cite{Jaeger1992}, the physics of granular materials depends on the complexities of inelastic interactions between neighboring particles \cite{Jaeger1996,Kadanoff1999,Aranson2006,Goldhirsch2003}. The influence of inelastic effects is perhaps most dramatically illustrated in the dynamics of a seemingly simple dilute gas of granular particles; even without any attractive interactions, dense particle clusters form due solely to energy loss during collisions \cite{Goldhirsch1993,McNamara1994,McNamara1996,Painter2003,Nichol2012}. These clusters subsequently collide and fragment. 
There has been much theory and simulation outlining the many different regimes for the dynamics in such gases.  However, there has been significantly less experimental work due to the difficulty of obtaining systems that do not immediately sediment due to gravity.
In this paper, we study one previously unappreciated regime, which can be studied by experiment as well as by theory and simulation, in which energy decays due to the collision of particle clusters.

Nearly all theory on granular gases have focused on the initial stages of cooling, where spatial density fluctuations are small and can be treated perturbatively \cite{Brilliantov2004}. Experiments have focused on this same regime \cite{Maass2008,Painter2003}. When three-body collisions are ignored, such theories cannot be applied to the dynamics of dense clusters. After clustering occurs, one dominant mode for cooling is from collisions between clusters. We show here that energy is rapidly dissipated during the initial moments of such a collision and the late-time behavior is determined by the early-time dynamics.

Upon collision, a cluster can be compressed so that the constituent particles become jammed and form many contacts that transmit force over long distances.  The static structure and linear response of jammed, mechanically-stable, packings has been well studied \cite{Liu2010}. In dynamics, experimental and theoretical work has shown the importance of ``jamming fronts" in dense granular systems. Most of these systems are influenced by external forces and/or are overdamped, such as clogging in confining flows \cite{Sheldon2010}, sedimentation of non-colloidal particles \cite{Davis1985}, or shear thickening in dense particle suspensions \cite{Waitukaitis2012}. These studies have identified two regions in the flow: one at low density where particles are not in contact, and one where the particles are jammed. The jamming front, whose speed and width depend on the details of the granular particles and the initial packing, is a boundary between the two regions \cite{Waitukaitis2013}. 

Here we report experiments and simulations showing how energy is dissipated by the growth of dynamic jamming fronts in the absence of external forces or constraints. Our experiment consists of the collision of two granular clusters composed of solid CO$_2$ (dry-ice) particles. These particles are levitated and float nearly frictionlessly on a hot surface so that the kinetic energy rapidly decays primarily due to inelastic collisions between particles. The initial loss of kinetic energy can be fit by: $\Delta E\equiv E(t)-E(t=0)=-Kt^{3/2}$, where $E(t)$ is the total kinetic energy at time $t$, and $t$ = 0 is the initial moment of impact. The prefactor $K$ depends on the velocity of the clusters, particle density, and the geometry of the overlapping region between the clusters. This form can be derived from a kinetic argument and can be generalized to other dimensions.

\begin{figure*}[!]
\begin{center}
\includegraphics[width=6.5 in]{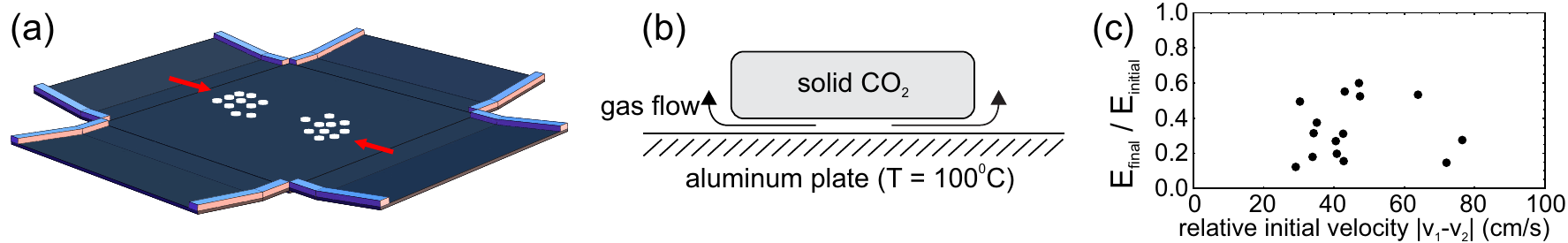}
\caption[]{(a) Schematic of the experimental apparatus. An anodized aluminum plate with tilted boundaries is heated to $\approx$ 100$^\circ$C. Two clusters of solid CO$_{2}$ particles impact in the middle of the plate. Silicone rubber strips prevent the particles from falling off the plate edges. The collision is filmed with a high-speed camera from above. (b) Sublimated gas from beneath a particle creates a high-pressure region which supports its weight. This leads to nearly frictionless translational motion. (c) Ratio of final to initial kinetic energy versus relative initial velocity for single-particle collisions.  There is a spread of values for the energy loss, even for similar relative velocities.
} 
\label{platefigure}
\end{center} 
\end{figure*}

\textit{Methods} --- The particles used in our experiments were cut from long rods of solid CO$_2$. Each nearly cylindrical particle had a radius of $\approx$ 0.8 cm and was $\approx$ 1.0 cm in length. The particles were levitated on a heated ($T$ $\approx$ 100$^\circ$C), cast aluminum plate of dimensions 61.0 cm $\times$ 61.0 cm $\times$ 1.25 cm by the Leidenfrost effect \cite{Quere2013,Burton2012}, where sublimation underneath the particles creates a supporting region of high pressure (Fig.\ \ref{platefigure}). The plate edges were slightly bent so that the particles sliding up and down them would conserve kinetic energy, mimicking elastic boundary conditions. The remaining edges were covered in silicone rubber to prevent particles from leaving the plate. In this geometry, the particles experience an essentially two-dimensional, force-free environment. 

We filmed the collisions from above using a high-speed digital camera (Phantom v9.0, Vision Research) with a resolution of 12.5 pixels/cm. The frame rate was 100 f/s with an exposure of 500 $\mu$s. To protect the aluminum surface and maximize visual contrast, the entire apparatus was anodized so that the dry-ice particles appeared white on a dark background. Experiments consisted of the collision between two clusters of $\approx$ 50--100 close-packed particles, which were initially at rest on the sloped boundaries and held in place with circular, plastic retainers. Upon removal of the retainer, the clusters gained momentum by sliding down the sloped boundaries, and subsequently collided near the middle of the plate. Particles at the rear slid further down the slope, thus gaining more momentum and causing a slight elongation of the clusters perpendicular to the direction of motion. The initial speed of the particles upon impact was $v_0$ $\approx$ 50 cm/s.

The fresh dry-ice particles were partially transparent, so that tracking their motion was complicated by inhomogeneities in pixel brightness. In addition, the particles were not always perfectly circular, so adjacent particles with flat edges prevented automated identification in some image frames. Thus, to measure particle velocities, we used a particle-image-velocimetry (PIV) method which correlates successive images to find displacements \cite{Burton2013}. Our algorithm was tested on images generated from a computer simulation of colliding clusters. Our PIV software is only sensitive to translational motion, so particle rotations were not measured. However, as we show below, our simulations indicate that rotations contribute insignificantly to the total kinetic energy. The mass of each particle was assumed to be proportional to their surface area in each image. There is no significant mass loss due to sublimation of the dry-ice particles over the duration of a cluster collision.

In order to characterize the individual collisions, we examined the impact of two isolated particles colliding at various velocities and impact angles. We measured the initial and final momentum, angular momentum, translational energy, and rotational energy by analyzing successive movie frames by hand. The final kinetic energy was 10--60\% of its initial value as shown in Fig.\ \ref{platefigure}c. Gas flow can propel small or irregularly-shaped particles in preferred directions \cite{Dupeux2013}. However, for the particle shape and size used in our experiments (circles, radius $\approx$ 0.8 cm), we found that the sublimated gas flow has an indiscernible effect on the particle trajectories.  

Our simulations use two-dimensional, time-integrated molecular dynamics following reference  \cite{Poschel2005}. The particles are monodisperse circles which interact via Hertzian elastic forces \cite{OHern2003}, viscous dissipation \cite{Brilliantov1996}, and tangential friction (coefficient $\mu$ = 0.5). In the simulations, all lengths are scaled by the particle radius, $\sigma$, all masses by the particle mass, $m$, and all times by $\sigma/c$, where $c$ is the speed of sound in an individual particle. We chose $v_0$ = 50/(3 $\times$ 10$^{5}$) = 1.6 $\times$ 10$^{-4}$ in the simulations to correspond to $c\approx$ 3 $\times$ 10$^{5}$ cm/s for our dry-ice particles.

For viscoelastic particles, the kinetic energy lost during collisions increases with impact velocity. We adjusted the ratio between the normal viscous and elastic forces so that the kinetic energy lost upon head-on impact of two particles was consistent with our measurements with dry-ice. In addition, we directly simulated the experiment by choosing initial positions and velocities from the first few frames of each experiment, and then compared the post-collision behavior of the experiment and simulation. The results were in excellent qualitative agreement. More details about the comparison between experiments and simulations can be found in reference \cite{Burton2013}.

\begin{figure}[!]
\begin{center}
\includegraphics[width=3.2 in]{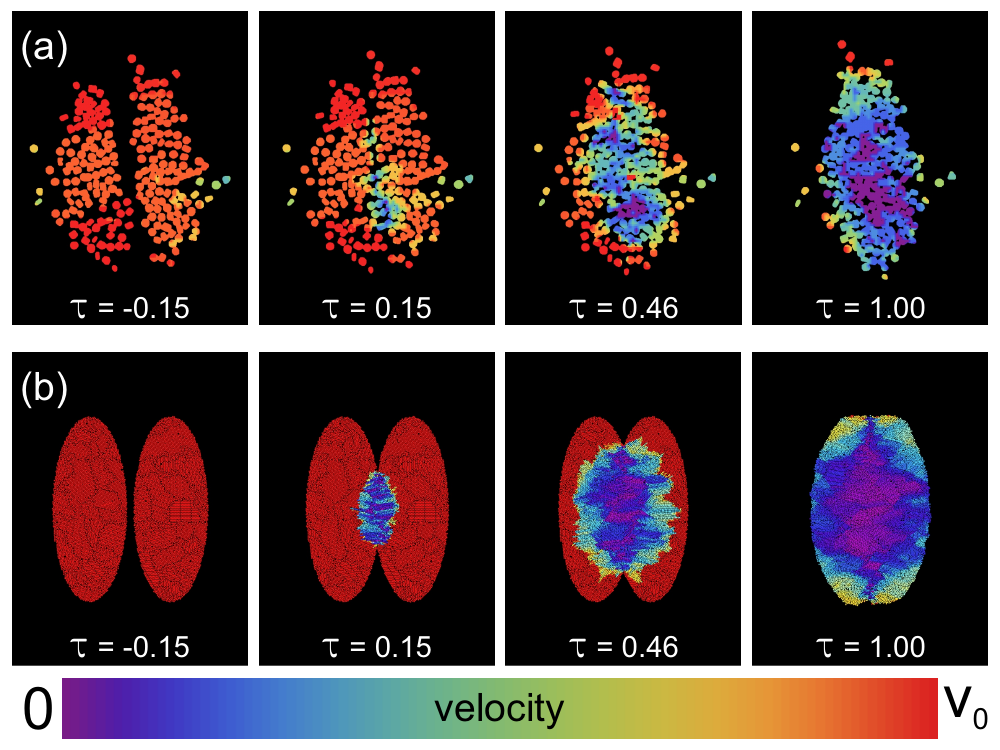}
\caption[]{(a) Images of the particles from an experiment during the initial moments of the collision. (b) Snapshots from a simulation of the collision between two elliptical clusters, each composed of 5000 particles. The initial area fraction inside each cluster is $\phi_0$ = 0.71. In both sets of images, the color indicates the magnitude of the velocity, as denoted by the scale bar on the bottom.  Upon impact, a jamming front spreads quickly and eventually encompasses all of the particles when $\tau\equiv t/t_{jam}$ = 1. 
} 
\label{velocity_color}
\end{center} 
\end{figure}

\textit{Growth of the jamming front} ---  In order to explore the dynamics of a cluster collision, we first look at the spatial distribution of particle velocities. Fig.\ \ref{velocity_color}a shows images from one experiment. Each particle is colored according to its velocity magnitude obtained from the PIV analysis. Initially, the particles are moving in the horizontal direction with constant velocity and appear red or orange. After the clusters collide, a region of reduced velocity appears (blue particles) and propagates throughout the cluster. In this region, particles are in close contact. As more particles collide from the rear, this jammed region grows and eventually encompasses the entire cluster. The time at which this occurs is denoted by $t_{jam}$, so that we may define a dimensionless time $\tau\equiv t/t_{jam}$. In our experiments, $t_{jam}\approx$ 0.13 s. After the collision, the resulting cluster of particles elongates in the vertical direction. In this regime, the energy decays much more slowly than during the initial impact as discussed in \cite{Burton2013}.

We compare these results with those obtained from our simulations. We start with two elliptical clusters (aspect ratio = 2.5), each composed of 5000 particles. The shape of the cluster was chosen to match the experiment, Fig.\ \ref{velocity_color}a, reasonably well.  (We have also simulated circular clusters and elliptical clusters with aspect ratios $<$ 1, and obtained similar results.) The area fraction of particles inside each cluster is $\phi_0$ = 0.71. Fig.\ \ref{velocity_color}b shows the initial moments of cluster impact. As in Fig.\ \ref{velocity_color}a, the red particles are traveling at impact velocity, $v_{0}$, and have not yet collided with any neighbors. The jamming front (indicated by the blue and purple particles) grows at a speed greater than $v_{0}$ until it encompasses both clusters. If we assume that the particles in the simulation are similar to dry-ice particles, then $t_{jam}\approx$ 0.5 s, which is larger than the experiment because the number of particles in the clusters are much larger in the simulation.

\begin{figure}[]
\begin{center}
\includegraphics[width=2.95 in]{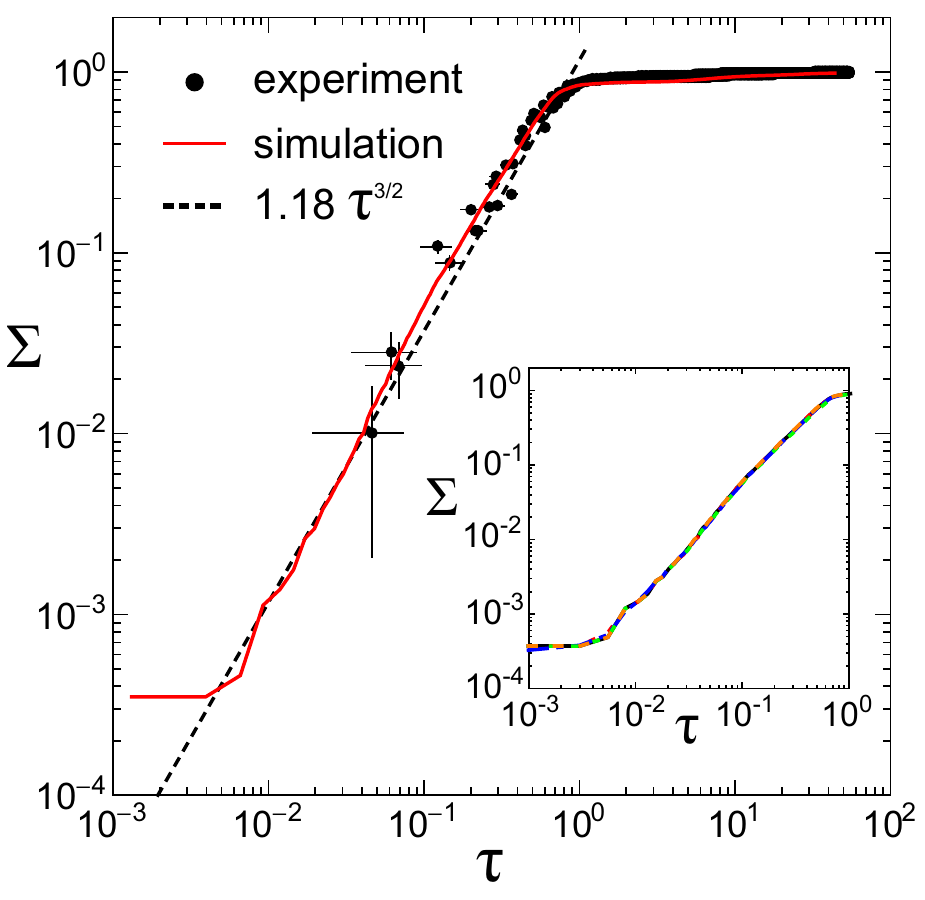}
\caption[]{Relative loss of kinetic energy, $\Sigma\equiv-\Delta E/E(t=0)$, versus $\tau\equiv t/t_{jam}$ immediately after impact for four experimental data sets (black dots) and the simulation of elliptical clusters (red line). The dashed line is the prediction (eqn. \ref{energyrat}) using parameters from the simulation. The inset compares five simulations using the same initial conditions, but with different dissipative forces.  Two have viscous forces that are 5$\times$ stronger and 5$\times$ weaker than in the main figure and two others have friction coefficients of $\mu$ = 0.1 and $\mu$ = 0.9. The results are virtually identical in all cases.
} 
\label{big_collision_energy}
\end{center} 
\end{figure}

\textit{Loss of kinetic energy} ---  We analyze the initial decay in kinetic energy after two clusters collide. In Fig.\ \ref{big_collision_energy} we plot $\Sigma\equiv-\Delta E/E(t=0)$, versus $\tau$ on a log-log plot. The data from the experiment (black points) are limited by the frame rate of the video and the number of particles. The red solid line, showing the simulation data, agrees well with the experiment over nearly two decades in $\Sigma$. At short times, the simulation data is flat because only two particles have collided. Eventually they collide with neighboring particles and the jamming front begins to spread, causing a further reduction in kinetic energy (i.e., increase in $\Sigma$).  We also measure the rotational kinetic energy in the simulations. We find that for all times shown in Fig.\ \ref{big_collision_energy}, rotations contribute less than 1\% to the total kinetic energy, so that $\Sigma$ is dominated by the loss of translational kinetic energy. 
 
To understand this behavior we generalize the one-dimensional, ``snowplow" model \cite{Waitukaitis2012,Waitukaitis2013} to higher dimensions. Qualitatively, the front velocity will increase with decreasing particle spacing. If, before a collision, there is space between the particles then this space must be traversed before particles collide with their neighbors; if the particles are already in contact, then the front will propagate near the speed of sound \cite{Gomez2012}. In two dimensions, we first compute the area of overlap of two colliding ellipses (with axis $a$ in the $x$-direction and axis $b$ in the $y$-direction), each moving at constant horizontal velocity, $v_0$. The total initial area of the ellipses is $A_0=2\pi a b$. As the ellipses first touch at $x=0$ and begin to overlap, the total area $A(t)$ is reduced:
\begin{align} 
A(t)=A_0-\frac{8b\sqrt{2}}{3\sqrt{a}}\delta^{3/2}+\mathcal{O}(\delta^{5/2}),
\label{areavst} 
\end{align}  
where $\delta=v_{0} t$ is the horizontal extent that each ellipse extends past $x$ = 0. The number of particles in both clusters is $\phi_{0}A_0$ where $\phi_0$ is the initial area fraction of particles in a cluster. To conserve particle number, the area lost in the overlapping region must be compensated by the increase in density in the jammed area, $A_{J}(t)$:
\begin{align} 
 \phi_{0} A_0= \phi_J A_J(t)+\phi_0 (A(t)-A_J(t)),
\label{areaeq} 
\end{align}  
where $\phi_J$ is the jammed area fraction. Solving for $A_J(t)$ to lowest order, we obtain
\begin{align} 
A_J(t)=\frac{8b\sqrt{2}}{3\sqrt{a}}\left(\frac{\phi_0}{\phi_J-\phi_0}\right)(v_0 t)^{3/2}.
\label{jammedarea} 
\end{align}  
If we assume that particles in the jammed region lose all their kinetic energy, then $\Sigma= (\phi_J A_J(t))/(\phi_0 A_0)$.  Combining equation \ref{areavst} with this assumption leads to a prediction for the initial decay in kinetic energy:
\begin{align} 
\Sigma=\frac{4 \sqrt{2}}{3 \pi}\left(\frac{\phi_J}{\phi_J-\phi_0}\right)\left(\frac{v_0 t}{a}\right)^{3/2}.
\label{energyrat} 
\end{align}  

We can compare this model with the simulation data in Fig.\ \ref{big_collision_energy}  by inserting the parameters used in the simulation ($t_{jam}$ = 1.87 $\times$ 10$^{5}$, $v_0$ = 1.6 $\times$ 10$^{-4}$, $a$ = 52.7, $\phi_0$ = 0.71), and assuming that $\phi_J=\pi/\sqrt{12}\approx0.907$ (the theoretical maximum for two-dimensional, monodisperse disks). Eqn.\ \ref{energyrat} then reduces to $\Sigma=1.18\tau^{3/2}$. This prediction, shown by the dashed line in Fig.\ \ref{big_collision_energy}, is in excellent agreement with the data.  Although eqn.\ \ref{energyrat} is specific for elliptical cluster shapes, the initial contact region between two arbitrarily-shaped clusters can always be quadratically expanded about the point of first contact. Thus, our results are generally valid at short times. Eqn.\ \ref{energyrat} can be generalized to other dimensions, $d$: $\Sigma\propto\tau^{(d+1)/2}$. Our results do not depend strongly on the particle restitution. The inset in Fig.\ \ref{big_collision_energy} shows four additional simulations with very different dissipative interactions compared with the simulation in the main figure. The results are insensitive to the dissipation parameters.

\textit{Conclusions} --- Dense granular clusters are a generic feature of freely-evolving granular systems and are commonly observed after an initial regime of cooling. Our experiments and simulations reveal the initial dynamics in the impact between two such clusters. A jamming front, whose velocity is much larger than the impact velocity, quickly spreads throughout the system. This produces a distinct regime for energy dissipation in the granular gas.  

Our results for the decay of the kinetic energy are valid at short times and can be computed using the density of the cluster and the geometry of the collision zone. The key assumption is that most of the energy is dissipated quickly. However, the cluster can also spread in the lateral direction \cite{Burton2013}. How spreading depends on the form of the dissipation is still an open question. Our studies have shown that the spreading occurs on a longer time scale, after most of the kinetic energy is dissipated.

Since dissipation can be accomplished through many rapid, slightly inelastic collisions or through a few highly inelastic ones, we expect that our conclusions should be valid as long as there are enough particles and collisions to dissipate the energy locally.  This recalls the results for shocks in inelastic gases \cite{BenNaim1999} and inelastic collapse \cite{McNamara1996}. Thus these results should be applicable to a wide range of natural and astronomical phenomena involving particles of varying material properties. 

We are grateful to Helmut Krebs for invaluable advice and assistance.  We thank Efi Efrati, Narayanan Menon and Scott Waitukaitis for important discussions. We acknowledge support from the Illinois Math and Science Academy (P.Y.L.), NSF DMR-1105145, MRSEC DMR-0820054 and PREM DMR-0934192 (J.C.B.) and the US Department of Energy, Office of Basic Energy Sciences, Division of Materials Sciences and Engineering, Award No. DE-FG02-03ER46088 (S.R.N.).

\end{document}